\documentclass[final,5p,times,twocolumn,number]{elsarticle}

\usepackage{amsmath,amssymb} 
\usepackage{graphicx}
\usepackage{xspace}
\usepackage{isotope}
\usepackage{orcidlink} 
\usepackage{lipsum}
\usepackage{siunitx}
\usepackage{upgreek}
\usepackage{booktabs}  
\usepackage{colortbl}
\usepackage{amsmath}
\usepackage[mathlines]{lineno} 
\usepackage[capitalise]{cleveref}
\usepackage{multirow}
\usepackage{microtype}
\usepackage{array}
\usepackage{dsfont}

\usepackage[normalem]{ulem}

\address[thuphy]{
Department of Physics,
Tsinghua University,
Beijing 100084, China
}

\address[thuhep]{
Center for High Energy Physics,
Tsinghua University,
Beijing 100084, China
}

\address[pkuphy]{
School of Physics and State Key Laboratory of Nuclear Physics and Technology,
Peking University,
Beijing 100871, China
}

\address[pkuhep]{
Center for High Energy Physics,
Peking University,
Beijing 100871, China
}

\address[columbia]{
Department of Physics,
Columbia University,
New York, NY 10027, USA
}

\address[cdutphy]{
College of Physics,
Chengdu University of Technology,
Chengdu 610059, China
}

\newcommand{\bea}{\begin{eqnarray}}
\newcommand{\eea}{\end{eqnarray}}

\begin{document}

\title{Cosmological Constrained Axion-Portal Inelastic Dark Matter for the LZ Event}

\author[thuphy,thuhep]{Haipeng An}
\ead{anhp@tsinghua.edu.cn}

\author[thuphy,thuhep]{Fei Gao}
\ead{feigao@tsinghua.edu.cn}

\author[pkuphy,pkuhep]{Jia Liu}
\ead{jialiu@pku.edu.cn}

\author[columbia]{Minghao Liu}
\ead{ml5107@columbia.edu}

\author[cdutphy]{Changlong Xu}
\ead{xuchanglong@cdut.edu.cn}

\begin{abstract}
The recent high-recoil candidate event reported by LUX-ZEPLIN
(LZ) motivates dark-matter scenarios with nonstandard kinematics
and momentum-dependent interactions. We study a two-state
inelastic dark-matter model in which $\chi_1$ and $\chi_2$ couple
off-diagonally to an axion-like particle (ALP) that also couples
to gluons and photons. Unlike treatments that assume only one
dark-matter state is present today, we track the cosmological
evolution of both states. The excited state $\chi_2$ is
sufficiently long-lived to survive to the present, with its
relic fraction determined by the dark-sector conversion process
$\chi_2\chi_2\leftrightarrow\chi_1\chi_1$.
We find that this fraction depends strongly on the Lorentz
structure of the DM--ALP interaction. Scalar transition
couplings lead to nonperturbative conversion rates with a
pronounced resonant structure, yielding a final excited-state
fraction that varies nonmonotonically with the mediator mass,
from $f_2\sim10^{-4}$ to approximately $1/2$ over the parameter
range studied. The relative importance of endothermic and
exothermic scattering is therefore parameter dependent.
By contrast, pseudoscalar transition couplings preserve
$f_2\simeq1/2$, yielding recoil spectra dominated by exothermic
$\chi_2 N\to\chi_1 N$ down-scattering for the benchmarks considered.
Benchmark spectra in all four scenarios can peak near the
observed recoil energy of $250\,\mathrm{keV}$. Our results
demonstrate that the dominant direction of inelastic scattering
in direct detection can be dynamically selected by the
early-Universe evolution of the dark sector. With additional
data, annual modulation measurements could distinguish these
scenarios. We further show that the ALP portal can be directly
probed at colliders.
\end{abstract}


\maketitle

\noindent{\it Introduction} --- The particle nature of dark matter (DM) remains one of the central open questions in particle physics and cosmology. Recently, the LUX-ZEPLIN (LZ) Collaboration reported a candidate event at a nuclear-recoil energy of $E_R=248\pm23,(\mathrm{stat})\pm23,(\mathrm{sys})~\mathrm{keV}$, with a maximum local significance of $3.4\sigma$ and a global significance of $2.6\sigma$~\cite{LZ:2026axp}. If interpreted as DM--nucleus scattering, such a high-energy recoil is difficult to accommodate with standard momentum-independent spin-independent elastic scattering, which favors lower recoil energies and is further suppressed by the xenon nuclear form factor in this energy range. This motivates inelastic kinematics, momentum-dependent interactions, or spin-dependent nuclear responses.

The theoretical interpretations of the LZ event can be broadly organized by the mechanism responsible for producing a high-energy recoil. The largest class relies on endothermic inelastic scattering, realized in Higgsino and other electroweak, scalar, $Z^\prime$, dark-photon, and related models~\cite{Fan:2026kxx,Freese:2026sga,Wu:2026nhi,Yin:2026jnn,Du:2026guj,Visinelli:2026kgt,Smirnov:2026aqk,Nomura:2026qyq,Su:2026rwz,DiMauro:2026ldr,Yamashita:2026ump,Lian:2026hpm,Ghosh:2026txe}. Exothermic or two-state scenarios~\cite{Baer:2026fpy,deLima:2026shq,Dent:2026bji,Paul:2026okh}, momentum- or spin-dependent elastic scattering~\cite{Unwin:2026rdp,Elahi:2026vlm,Khan:2026nwp}, boosted DM~\cite{Alhazmi:2026efz,Kannike:2026qyl,Liang:2026coz,Heikinheimo:2026kwp}, fermionic DM absorption~\cite{Lou:2026idn}, and alternative neutrino or nuclear explanations~\cite{Jeesun:2026vzo,Chattaraj:2026fxn,Aghaie:2026vsu,Gu:2026vto} have also been considered. Several subsequent works have examined complementary constraints and tests of these interpretations~\cite{Pospelov:2026ewn,Rodd:2026tyn,Kotlarski:2026pep,Mahapatra:2026glu}.

In this work, we revisit the two-state inelastic-DM framework~\cite{An:2020tcg} in the context of the LZ high-recoil event, with the two states $\chi_{1,2}$ coupled off diagonally to an axion-like particle (ALP) of mass $m_a\sim\mathcal O(0.1\text{--}10)\mathrm{GeV}$. The ALP couples to gluons and photons through $aG\widetilde G$ and $aF\widetilde F$, with the gluonic pseudoscalar interaction generating a momentum-dependent, spin-dependent nuclear response that is well suited to the high-recoil regime. The secluded annihilation channel $\chi\chi\to aa$, followed predominantly by ALP decays into hadrons or photons, also avoids the strong solar-neutrino constraints that apply to several electroweak inelastic-DM interpretations. ALP-mediated endothermic scattering has recently been considered for the LZ event~\cite{Yuan:2026djt}, but here we instead track the cosmological evolution of both DM states. We find that, for portal couplings relevant to the LZ signal, the decay $\chi_2\to\chi_1\gamma\gamma$ is sufficiently slow that $\chi_2$ can survive to the present day. Its abundance is then controlled by dark-sector conversion processes such as $\chi_2\chi_2\leftrightarrow\chi_1\chi_1$. Remarkably, the resulting excited-state fraction depends strongly on the Lorentz structure of the DM--ALP interaction:
scalar transition couplings yield an oscillatory excited-state fraction spanning \(f_2\sim10^{-4}\) to \(1/2\), whereas pseudoscalar transition couplings leave \(f_2\simeq1/2\).
Consequently, both endothermic $\chi_1 N\to\chi_2 N$ and exothermic $\chi_2 N\to\chi_1 N$ scattering can contribute to the LZ recoil spectrum, directly linking the signal to the cosmological evolution of the two-state dark sector.

\noindent{\it The model} --- We consider two nearly degenerate fermionic DM states, $\chi_1$ and $\chi_2$, interacting through an axion-like particle (ALP) $a$. The relevant interactions are
\begin{align}
\mathcal L \supset {}&
y a \overline{\chi}_1\Gamma\chi_2+\mathrm{h.c.}
\\
&+
\frac{C_G}{f_a}\frac{\alpha_s}{8\pi}
aG_{\mu\nu}^{A}\widetilde G^{A\mu\nu}
+
\frac{C_F}{f_a}\frac{\alpha_{\rm em}}{8\pi}
aF_{\mu\nu}\widetilde F^{\mu\nu}. \nonumber
\label{eq:model}
\end{align}
Here $C_G$ and $C_F$ are dimensionless portal couplings and $f_a$ denotes the ALP scale. We take $\chi_1$ to be the lighter state and define the positive mass splitting $\Delta\equiv m_2-m_1>0$, with $m_1\simeq m_2\equiv m_{\rm DM}$ and $\Delta=\mathcal O(300)\,\mathrm{keV}$. We focus on $m_a\sim\mathcal O(0.1\text{--}10) \,\mathrm{GeV}\ll m_{\rm DM}$. The gluonic coupling mediates DM--nucleus scattering, while the photon coupling induces $\chi_2\to\chi_1\gamma\gamma$ and therefore controls the decay lifetime of the excited state.

We consider four benchmark scenarios, distinguished by the fermionic nature of $\chi_{1,2}$ and the Lorentz structure of their transition coupling to the ALP. In S1 and S2, $\chi_{1,2}$ are Majorana fermions, with $\Gamma=\mathds{1}/2$ and $i\gamma_5/2$, respectively, where the conventional factor of $1/2$ is included in the Majorana interaction Lagrangian to avoid double counting. In S3 and S4, $\chi_{1,2}$ are Dirac fermions, with $\Gamma=\mathds{1}$ and $i\gamma_5$. The four scenarios, together with the coupling required to reproduce the observed thermal relic abundance in the limit $m_a\ll m_{\rm DM}$, are summarized in Table~\ref{tab:dm_scenarios}. The discussions of the UV completion model building is presented in~\ref{app:UV}. 

\begin{table}[t]
\centering
\caption{Benchmark scenarios and the approximate coupling required by the observed DM relic abundance, for $m_a\ll m_{\rm DM}$.}
\label{tab:dm_scenarios}
\setlength{\tabcolsep}{3.5pt}
\renewcommand{\arraystretch}{1.25}
\begin{tabular}{c c c c}
\hline\hline
Scenario & DM type & $\Gamma$ & $y_{\rm relic}$ \\
\hline
S1 & Majorana & $\mathds{1}/2$
& $\displaystyle 1.4\sqrt{m_{\rm DM}/{\rm TeV}}$ \\
S2 & Majorana & $i\gamma_5/2$
& $\displaystyle 2.5\sqrt{m_{\rm DM}/{\rm TeV}}$ \\
S3 & Dirac & $\mathds{1}$
& $\displaystyle 1.7\sqrt{m_{\rm DM}/{\rm TeV}}$ \\
S4 & Dirac & $i\gamma_5$
& $\displaystyle 3.0\sqrt{m_{\rm DM}/{\rm TeV}}$ \\
\hline\hline
\end{tabular}
\end{table}

\bigskip

\noindent{\it DM freeze-out} --- For $m_a\ll m_{\rm DM}$ and $\Delta\ll T_{\rm fo}$, the two DM states are effectively degenerate during chemical freeze-out, and the total relic abundance is set predominantly by annihilation into ALP pairs. The relevant processes are $\chi_i\chi_i\to aa$ for Majorana DM and $\chi_i\bar\chi_i\to aa$ for Dirac DM, with $i=1,2$. Averaging over the nearly degenerate states gives
\begin{equation}
\langle\sigma v\rangle_{\rm eff}=
\begin{cases}
\dfrac{1}{2}\langle\sigma v\rangle_{ii\to aa}, & \text{Majorana}, \\
\dfrac{1}{4}\langle\sigma v\rangle_{i\bar i\to aa}, & \text{Dirac}.
\end{cases}
\label{eq:sigmav_eff}
\end{equation}
In the limit $m_a\ll m_{\rm DM}$, these annihilation cross sections depend only weakly on $m_a$. Requiring the observed DM relic abundance therefore fixes $y$ approximately as a function of $m_{\rm DM}$. The resulting relations are given in Table~\ref{tab:dm_scenarios}. The subsequent redistribution between $\chi_1$ and $\chi_2$ is governed by dark-sector conversion and will be discussed below.
\\

\noindent{\it Lifetime of $\chi_2$}---For $m_a\gg\Delta$, the ALP can be integrated out, yielding
\begin{equation}
{\cal L}_{\rm eff}=
\frac{\alpha_{\rm em}}{8\pi}\,
\widetilde C_F\,
F_{\mu\nu}\widetilde F^{\mu\nu}
\bar\chi_1\Gamma\chi_2
+{\rm h.c.},
~
\widetilde C_F\equiv\frac{yC_F}{m_a^2f_a}.
\end{equation}
This operator induces the three-body decay
$\chi_2\to\chi_1\gamma\gamma$. In the limit
$\Delta\ll m_a\ll m_{\rm DM}$, the corresponding lifetimes are
\begin{align}
& \tau_{\rm S1,S3} =
\frac{3360\pi^5}
{\alpha_{\rm em}^2\widetilde C_F^2\Delta^7} \\
&\simeq
1.3\times10^{12}\ {\rm yr}
\left[
\frac{(30~{\rm GeV})^{-3}}{\widetilde C_F}
\right]^2
\left(\frac{300~{\rm keV}}{\Delta}\right)^7 ,
\label{eq:lifetime} \nonumber \\
&\tau_{\rm S2,S4}
=
\frac{40320\pi^5m_{\rm DM}^2}
{\alpha_{\rm em}^2\widetilde C_F^2\Delta^9}
\\
&\simeq
2.5\times10^{17}\ {\rm yr}
\left[
\frac{(1~{\rm GeV})^{-3}}{\widetilde C_F}
\right]^2
\left(\frac{300~{\rm keV}}{\Delta}\right)^9
\left(\frac{m_{\rm DM}}{1~{\rm TeV}}\right)^2 . \nonumber
\end{align}
The benchmark values of $\widetilde C_F$ are chosen to be representative of the corresponding gluonic coupling
$\widetilde C_G\equiv yC_G/(m_a^2f_a)$ required by the LZ recoil rate. Even for $C_F=C_G$, the resulting lifetime is well above the age of the Universe in all four scenarios. Decays of $\chi_2$ therefore do not appreciably deplete its cosmological abundance; its present-day fraction is instead determined by dark-sector conversion processes.

The decay $\chi_2\to\chi_1\gamma\gamma$ also injects electromagnetic energy into the cosmological plasma and produces X-ray/soft-$\gamma$-ray photons in the Galactic halo. Since $m_a\gg\Delta$, this is a genuine three-body decay with a continuous photon spectrum and total photon energy $E_{\gamma_1}+E_{\gamma_2}\simeq\Delta$. Relative to conventional DM decay $X\to\gamma\gamma$ with $m_X\sim\Delta$, both the cosmological energy injection and the Galactic photon flux are suppressed by
\begin{equation}
\epsilon_{\rm inj}
\simeq
f_2\frac{\Delta}{m_{\rm DM}},
\end{equation}
where $f_2$ is the excited-state fraction. Thus, bounds on $X\to\gamma\gamma$ can be approximately recast as
\begin{equation}
\tau_{\chi_2}
\gtrsim
f_2\frac{\Delta}{m_{\rm DM}}\,
\tau_{\gamma\gamma}^{\rm bound}\quad (m_X\sim\Delta),
\end{equation}
up to corrections associated with the continuous three-body spectrum~\cite{Slatyer:2016qyl,Liu:2020wqz,Capozzi:2023xie,Liu:2023nct}. 
For $\Delta\simeq300~{\rm keV}$ and $m_{\rm DM}\simeq1~{\rm TeV}$, we have $\epsilon_{\rm inj} \simeq 1.5\times10^{-7}$ even for the naive estimate $f_2\simeq1/2$, which provides strong suppression renders the CMB and IGM-heating limits very weak. 

Galactic X-ray observations provide a more direct probe of the same decay. In particular, Ref.~\cite{Krnjaic:2025zjl} derived limits on the continuous photon spectrum from $\chi_2\to\chi_1\gamma\gamma$ using 16 years of INTEGRAL/SPI data. For $\Delta\simeq300~{\rm keV}$, their scalar-mediated result gives approximately
\begin{equation}
\tau_{\chi_2}m_{\rm DM}
\gtrsim
1.2\times10^{24}~{\rm s\,GeV}
\left(\frac{f_2}{1/2}\right),
\end{equation}
where the published limit assumes $f_2=1/2$. For $m_{\rm DM}=1~{\rm TeV}$, this corresponds to
$\tau_{\chi_2}\gtrsim3.8\times10^{13}~{\rm yr}\,(f_2/1/2)$. This bound is immediately satisfied for S2 and S4, while for S1 and S3 the smaller excited-state abundance derived below substantially weakens the constraint.
\\

\noindent{\it Excited-state depletion}---
At early times, bath-assisted transitions such as
$\chi_1 g\leftrightarrow\chi_2 g$ and
$\chi_1\gamma\leftrightarrow\chi_2\gamma$ maintain chemical
equilibrium between the two states, giving
$n_2/n_1\simeq e^{-\Delta/T}$ and hence $f_2\simeq1/2$ for
$T\gg\Delta$.  These processes decouple before dark-sector
de-excitation becomes important.  After kinetic decoupling at
$T_d\simeq\min(\Lambda_{\rm QCD},m_a)$, the DM temperature redshifts
approximately as
\begin{equation}
T_\chi\simeq\frac{T^2}{T_d}.
\label{eq:Tchi}
\end{equation}
When $T_\chi\sim\Delta$, corresponding to
$T\sim\sqrt{\Delta T_d}\sim{\cal O}(5)\,{\rm MeV}$ for our
benchmarks, the photon-induced conversion rate is already well below
the Hubble rate.  The subsequent evolution of the excited-state
abundance is therefore governed by
$\chi_2\chi_2\leftrightarrow\chi_1\chi_1$.

Defining
\begin{equation}
f_2\equiv\frac{n_2}{n_1+n_2},
\qquad
x\equiv\frac{\Delta}{T_\chi},
\end{equation}
the Boltzmann equation becomes
\begin{equation}
\frac{df_2}{dx}
=
-Ax^{-3/2}
\left[
f_2^2-(1-f_2)^2e^{-2x}
\right],
\label{eq:df2}
\end{equation}
where
\begin{equation}
A=
\frac{\Omega_{\rm DM}\rho_c}{m_\chi s_0}
\frac{2\pi^2}{45}
\frac{g_{*s}}{2\times1.66\sqrt{g_*}}
M_{\rm Pl}\sqrt{\Delta T_d}\,
\langle\sigma v\rangle_{22\to11}.
\label{eq:A}
\end{equation}
The derivation of Eq.~\eqref{eq:df2}, including the Dirac
particle--antiparticle bookkeeping, is given in the Supplemental
Material.

An important feature of the exothermic conversion is that the mass
splitting itself generates a finite momentum transfer even for
negligible initial kinetic energy,
\begin{equation}
q_*^2\simeq2m_{\rm DM}\Delta,
\qquad
\frac{1}{t-m_a^2}\simeq
-\frac{1}{m_a^2+2m_{\rm DM}\Delta}.
\label{eq:qstar}
\end{equation}
For $m_{\rm DM}\sim1~{\rm TeV}$ and
$\Delta\sim300~{\rm keV}$, one finds $q_*\simeq0.8~{\rm GeV}$.
The momentum dependence of the mediator propagator therefore cannot
be neglected.

In the regime $T_\chi \ll \Delta$, the conversion cross sections can be
evaluated in the nonrelativistic limit. For scenario S2, a perturbative
calculation yields
\begin{equation}
\left\langle \sigma v \right\rangle_{22\to11}^{\mathrm{S2}}
\simeq
\frac{y^4}{8\sqrt{2}\pi}\,
\frac{\Delta^{5/2}}
{m_\chi^{1/2}\left(m_a^2+2m_\chi\Delta\right)^2}.
\label{eq:sigmav}
\end{equation}
The corresponding Dirac scenario, S4, has the same parametric dependence,
but the cross section is larger by an overall factor of $3/2$.

\begin{figure}[htpb]
    \centering
    \includegraphics[width=\columnwidth]{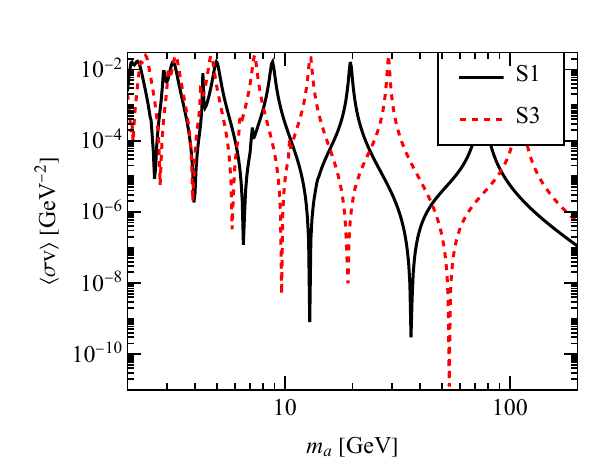}
    \includegraphics[width=\columnwidth]{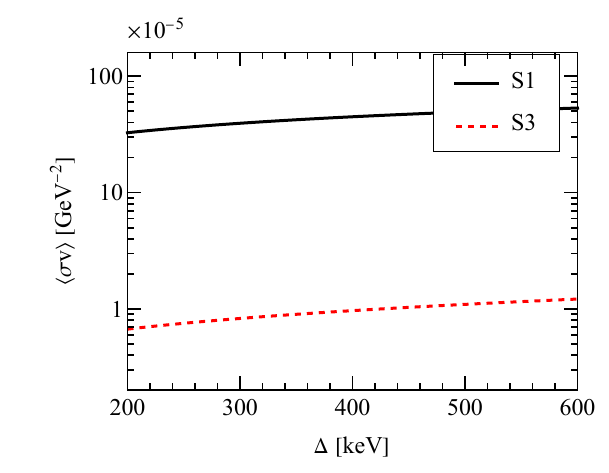}
    \caption{Numerical results for $\langle\sigma v\rangle_{22\rightarrow11}$ for S1 (black solid) and S3 (red dashed), with $m_\chi=900\,\mathrm{GeV}$ and $\Delta/T=20$. Top: dependence on $m_a$ at fixed $\Delta=200\,\mathrm{keV}$. Bottom: dependence on $\Delta$ at fixed $m_a=10\,\mathrm{GeV}$.}
    \label{fig:S13}
\end{figure}

\begin{figure}[ht!]
\centering
\includegraphics[width=0.5\textwidth]{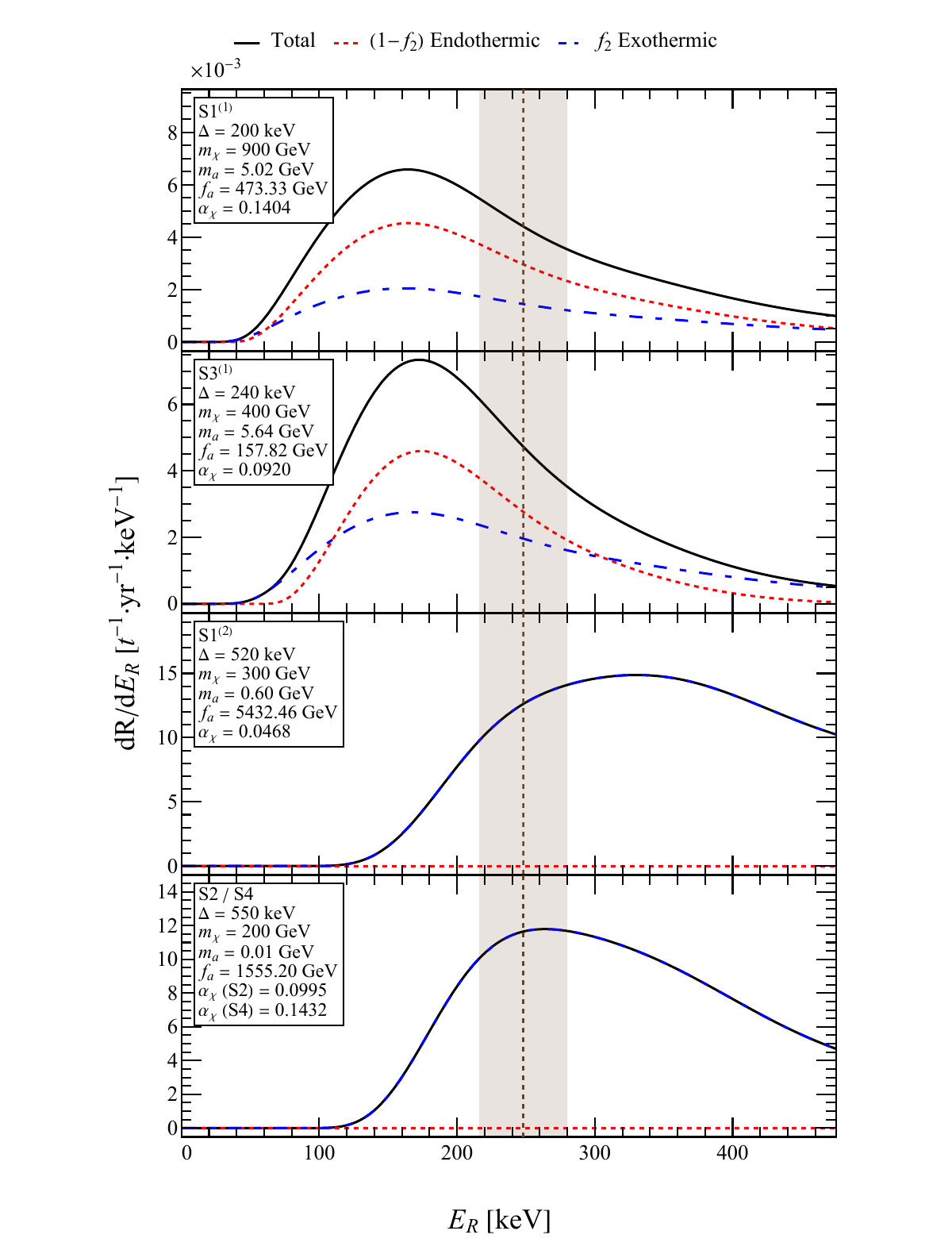}
\caption{Differential nuclear recoil spectra for the benchmark
points. The upper three panels correspond to S1/S3, where
endothermic scattering dominates, while the lower panel
corresponds to S2/S4, where exothermic scattering dominates.
The vertical dashed line marks the recoil energy of the LZ
event, and the shaded band indicates the associated energy
uncertainty.}
\label{fig:spec}
\end{figure}

In scenarios S1 and S3, the axion $a$ couples to a scalar bilinear of $\chi_1$ and $\chi_2$. Consequently, the conversion process $\chi_2\chi_2 \to \chi_1\chi_1$ is not velocity suppressed. Nonperturbative effects are expected to become important when
\begin{equation}
\alpha_\chi m_\chi \gg m_a,
\qquad
v_{\mathrm{rel}} \lesssim \alpha_\chi,
\end{equation}
where $v_{\mathrm{rel}}$ is the relative velocity of the incoming particles. As shown in the legends in the two upper panels of Fig.~\ref{fig:spec}, both conditions are satisfied by the S1 and S3 benchmarks that account for the LZ event. The Born approximation is therefore inadequate for these benchmarks, and determining the excited-state fraction $f_2$ requires explicitly solving the coupled-channel Schr\"odinger equation. The procedure for calculating $f_2$ in S1 and S3 is outlined in~\ref{app:nonpert_conversion}. Fig.~\ref{fig:S13} shows $\langle\sigma v\rangle_{22\rightarrow11}$ as a function of $m_a$ and $\Delta$ for representative parameter choices in S1 and S3, revealing a pronounced resonant structure induced by the long-range interaction.

\begin{figure}[htpb]
    \centering
    \includegraphics[width=\columnwidth]{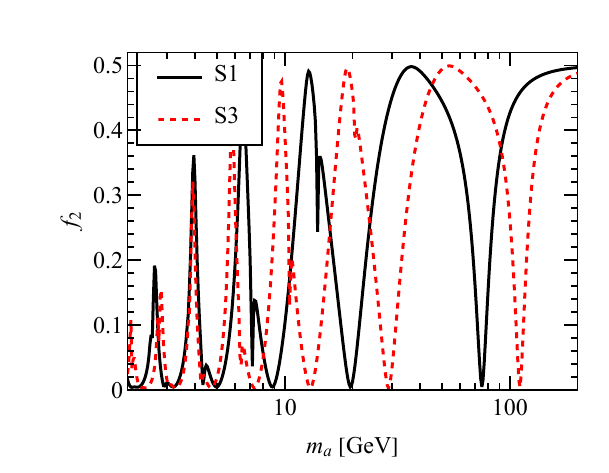}
    \caption{The excited-state fraction $f_2$ as a function of the ALP mass $m_a$, obtained by solving the Boltzmann equation for $m_\chi=900\,\mathrm{GeV}$ and $\Delta=200\,\mathrm{keV}$. The Yukawa coupling $y$ is fixed by the DM relic abundance.}
    \label{fig:f2B}
\end{figure}

The finite-range interaction mediated by the massive axion can produce
resonant structure in the conversion rates
$(\sigma v)_{22\to11}^{\mathrm{S1,S3}}$.
Accordingly, we evaluate $f_2$ numerically rather than using a
closed-form expression. For S1 and S3 the result of $f_2$ as a function of $m_a$ is shown in Fig.~\ref{fig:f2B}


\begin{figure}[t]
    \centering
    \includegraphics[width=\columnwidth]{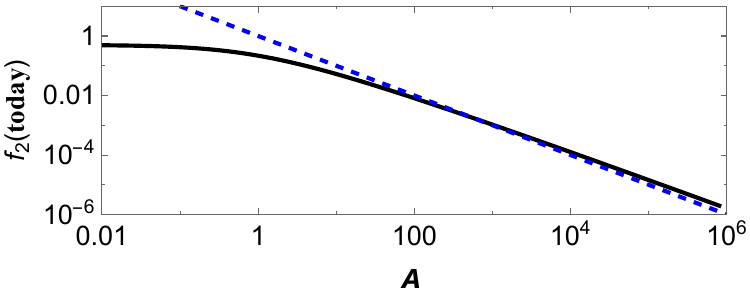}
    \caption{Present-day excited-state fraction $f_2$ as a function of the
dimensionless conversion parameter $A$. The black curve is
obtained by solving Eq.~\eqref{eq:df2}. For $A \ll 1$,
$f_2(\mathrm{today}) \approx 1/2$, while for $A \gg 1$,
$f_2(\mathrm{today}) \approx A^{-1}$, as indicated by the
blue dashed line.}
    \label{fig:f2A}
\end{figure}

For S2 and S4, using the values of $y$ fixed by the relic abundance, the value of $A$ can be obtained analytically
\begin{equation}
\begin{aligned}
A_{\rm S4}\simeq 3A_{\rm S2}
&\simeq
6.51\times10^{-6}\,\mathcal F
\left(\frac{m_\chi}{1\,{\rm TeV}}\right)^{1/2}
\left(\frac{\Delta}{300\,{\rm keV}}\right)^3,
\end{aligned}
\label{eq:AS}
\end{equation}
where
\begin{equation}
\mathcal F=
\left(\frac{m_a}{1\,{\rm GeV}}\right)^{-7/2}
\sqrt{\frac{T_d}{m_a}}\,
\left(1+\frac{2m_\chi\Delta}{m_a^2}\right)^{-2}.
\end{equation}
Figure~\ref{fig:f2A} shows the present-day excited-state fraction
obtained by numerically solving Eq.~\eqref{eq:df2}.  For $A\ll1$,
depletion is negligible and $f_2({\rm today})\simeq1/2$, whereas for
$A\gg1$ the excited-state abundance decreases approximately as
$f_2\propto A^{-1}$.  In the parameter region relevant to the LZ
event, S2 and S4 have inefficient conversion and retain
$f_2({\rm today})\simeq1/2$.  By contrast, S1 and S3 typically  yield
$f_2({\rm today})\sim O(10^{-4})$--$O(1/2)$.
Consequently, endothermic and exothermic scattering can make comparable contributions to the LZ recoil spectrum in S1 and S3, whereas exothermic scattering dominates in S2 and S4.

\bigskip

\noindent{\it Nuclear recoil spectrum} --- To calculate the nuclear recoil rate and spectrum, we follow
Ref.~\cite{An:2025bby}, using \texttt{DirectDM}~\cite{Bishara:2017nnn}
to match the partonic axion--gluon interaction onto low-energy
nucleon operators and \texttt{DMFormFactor}~\cite{Anand_2014} to
evaluate the corresponding nuclear responses and recoil spectra.
As shown below, the benchmark points that are consistent with
the cosmological evolution and can fit the LZ data have mediator
masses in the range $m_a \simeq 0.01$--$10\,\mathrm{GeV}$. Since
these masses are not necessarily large compared with the relevant
momentum transfer, we retain the full momentum dependence of
the $t$-channel axion propagator. In the laboratory frame, where
the initial nucleus is at rest, the propagator takes the form
\begin{equation}
    \frac{1}{(p_2-p_1)^2-m_a^2}
    =
    -\frac{1}{2m_A E_R+m_a^2},
\end{equation}
where $p_2$ and $p_1$ are the incoming and outgoing dark-matter
four-momenta, respectively, $m_A$ is the target nuclear mass,
and $E_R$ is the nuclear recoil energy. This expression assumes
that the nucleus remains in its ground state.

In the CP-violating scenarios S1 and S3, the axion couples to the
scalar transition bilinear $\bar{\chi}_1\chi_2$. At leading order
in the nonrelativistic expansion, and after factoring out the
external-state normalization, this bilinear matches onto the
identity operator in dark-matter spin space, $\mathds{1}_\chi$,
for both Majorana and Dirac fermions. Consequently, S1 and S3
yield identical recoil spectra for the same masses, physical
transition couplings, and incident dark-matter flux. Similarly,
in S2 and S4, the axion couples to the pseudoscalar transition
bilinear $\bar{\chi}_1 i\gamma_5\chi_2$, whose leading
nonrelativistic reduction is proportional to
$\vec{q}\cdot\vec{S}_\chi$, up to a convention-dependent phase.
Here, $\vec{q}$ is the three-momentum transfer and $\vec{S}_\chi$
is the dark-matter spin operator. S2 and S4 therefore also yield
identical recoil spectra under the same conditions.

In Fig.~\ref{fig:spec}, we show the nuclear recoil spectra for
three benchmark points: three for S1/S3 and one for S2/S4. We choose
$\Delta$ and $m_\chi$ such that each spectrum peaks near the recoil
energy of the observed LZ event. The parameters $m_a$ and $f_a$
are chosen to yield one expected event over the LZ exposure,
while ensuring that the axion--gluon coupling remains consistent
with current heavy-axion search constraints
(see Fig.~\ref{fig:constraints} in the Appendix).
For the S1/S3 benchmarks in the upper two panels, the parameters
are chosen such that endothermic and exothermic scattering make
comparable contributions, with $f_2\sim\mathcal{O}(10^{-4})$.
Despite the small excited-state population, exothermic scattering
remains important because endothermic up-scattering probes the
suppressed high-velocity tail of the halo distribution.
When $f_2\simeq1/2$, exothermic down-scattering dominates,
as illustrated by benchmark points $\rm S1^{(2)}$ and S2/S4 in the third and fourth panel.

\bigskip

\noindent{\it Annual modulation} --- The Earth's orbital motion around the Sun induces a seasonal
variation in the dark-matter velocity distribution in the
laboratory frame, giving rise to an annual modulation of the
event rate. 
We model the dark-matter halo with a truncated Maxwellian velocity distribution, characterized by a most probable speed $v_0=220~{\rm km/s}$ and an escape velocity $v_{\rm esc}=550~{\rm km/s}$. The Earth's speed relative to the dark-matter
halo is parametrized as
\begin{equation}
v_E(t)
=
232~{\rm km/s}
+
15~{\rm km/s}\times\cos\left[\omega(t-t_0)\right],
\end{equation}
where $\omega=2\pi/{\rm yr}$ and $t_0$ denotes the time at which
the Earth's speed reaches its maximum.

\begin{figure}[htpb]
    \centering
    \includegraphics[width=0.9\columnwidth]{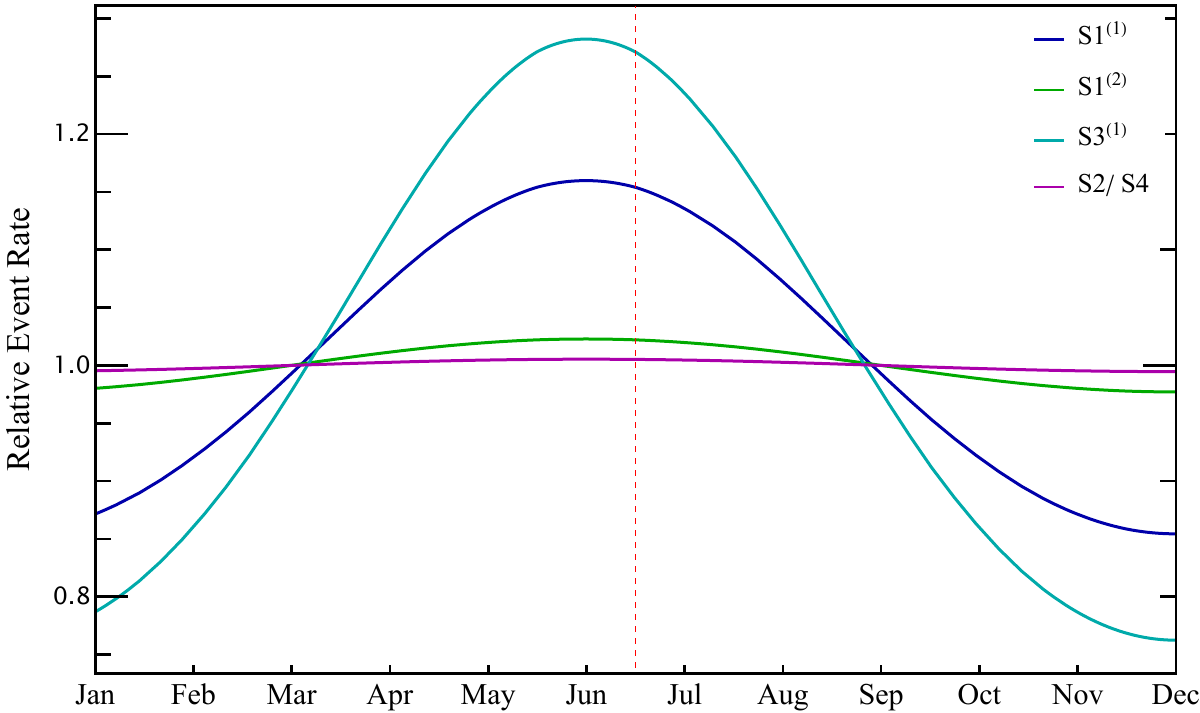}
    \caption{Predicted temporal evolution of the signal rate for the four benchmark
points. The relative event rate is normalized to unity for a
time-independent signal rate. The vertical dashed line marks the
timestamp of the LZ event.}
    \label{fig:modulation}
\end{figure}

In scenarios S2/S4, nuclear recoils are dominated
by exothermic down-scattering, in which the recoil energy receives
contributions from both the mass splitting $\Delta$ and the
kinetic energy of the incoming $\chi_2$. The energy released
by the transition reduces the sensitivity to seasonal variations
in the incident kinetic energy, leading to a relatively small
modulation for these benchmarks. As shown in Fig.~\ref{fig:modulation}, the annual modulation can
be safely neglected for the benchmark points $\rm S2/S4$.

For endothermic scattering, the minimum incoming $\chi_1$ speed
required to produce a nuclear recoil of energy $E_R$ is
\begin{equation}\label{eq:vmin}
    v_{\rm min}(E_R)
    =
    \frac{1}{\sqrt{2m_N E_R}}
    \left| \frac{m_N E_R}{\mu_N} + \Delta \right|,
\end{equation}
where $m_N$ is the target nuclear mass and $\mu_N$ is the reduced
mass of the $\chi_1$--nucleus system. For the benchmark considered
here, $E_R \sim 250\,\mathrm{keV}$ corresponds to
$v_{\rm min} \approx 600\,\mathrm{km\,s^{-1}}$, placing the
scattering threshold in the high-velocity tail of the
laboratory-frame dark-matter distribution. In this regime,
the seasonal variation in the Earth's speed relative to the
halo, of order $15$--$20\,\mathrm{km\,s^{-1}}$, can appreciably
alter the fraction of particles above the scattering threshold.
The endothermic-dominated recoil rate is therefore expected to
exhibit an enhanced fractional annual modulation. Indeed,
Fig.~\ref{fig:modulation} shows that the rate peaks in June,
significantly exceeding its annual average.

The modulation amplitude depends on both the mass splitting
$\Delta$, which sets the minimum speed through
Eq.~\eqref{eq:vmin}, and the relative contributions of
endothermic and exothermic scattering. As shown in
Fig.~\ref{fig:modulation}, the fractional modulation amplitude
reaches approximately $15\%$ for S1$^{(1)}$ and exceeds $20\%$
for S3$^{(1)}$. These benchmarks exhibit appreciable seasonal
variations, with the signal rate peaking near June, close to
the June 16 observation date of the LZ candidate event.
By contrast, S1$^{(2)}$ and S2/S4 exhibit negligible annual
modulation. These differences illustrate how timing information
could help distinguish benchmark scenarios with similar
recoil-energy spectra.


\bigskip

\noindent{\it Conclusions and discussions} ---
We have investigated the recent LZ high-recoil candidate event
within a two-state inelastic-DM model coupled to the SM through
an ALP portal. The gluonic coupling induces a momentum-dependent,
spin-dependent nuclear interaction, while the photon coupling
enables the excited-state decay $\chi_2\to\chi_1\gamma\gamma$.
In the parameter region relevant to the LZ event, this decay is
sufficiently slow for the excited state to survive to the present.
Its present-day abundance is instead set by the dark-sector
conversion process $\chi_2\chi_2\leftrightarrow\chi_1\chi_1$.
We find a strong dependence on the Lorentz structure of the
DM--ALP interaction:  in the scalar-transition scenarios S1 and
S3, nonperturbative conversion yields an excited-state fraction
with an oscillatory dependence on the mediator mass, ranging
from $f_2\sim10^{-4}$ to approximately $1/2$ over the parameter
range studied. Endothermic $\chi_1 N\to\chi_2 N$ and exothermic
$\chi_2 N\to\chi_1 N$ scattering can make comparable contributions
even for $f_2\sim10^{-4}$, whereas exothermic scattering dominates
when $f_2\simeq1/2$. By contrast,
the pseudoscalar-transition scenarios S2 and S4 retain
$f_2\simeq1/2$, leading to spectra dominated by exothermic
$\chi_2 N\to\chi_1 N$ down-scattering. In all four scenarios,
benchmark spectra can peak near the observed recoil energy
of $250\,\mathrm{keV}$.

This connection between early-Universe dark-sector dynamics
and the dominant direction of inelastic scattering can be
tested with xenon-detector data. The recently reported exposures
of LZ, PandaX-4T, and XENONnT are approximately $5.7$, $1.5$,
and $6.8$ tonne-years, respectively
\cite{LZ:2026cevns,PandaX:2024wimp,XENON:2026cevns}.
If the LZ candidate originates from DM and the event yield
scales approximately with exposure, the combined data could
already contain of order five such events. Their recoil-energy
and temporal distributions could help test the DM interpretation
and, with sufficient statistics, distinguish endothermic from
exothermic scattering, thereby probing the cosmological
population of the excited DM state. Furthermore, current and future particle physics experiments, including the High-Luminosity LHC~\cite{Hook:2019qoh}, DUNE~\cite{Kelly:2020dda}, FASER~\cite{FASER:2018eoc}, NA62~\cite{Ertas:2020xcc}, and SHIFT~\cite{Niedziela:2024khw} can provide complementary probes of the ALP portal (see \ref{sec:appendixB} for a detailed discussion).

In scenarios S1 and S3, $\chi_1$ and $\chi_2$ can also form bound states, potentially enhancing the effective annihilation rate and substantially altering the viable parameter space. A detailed treatment of these effects is beyond the scope of this work. We therefore restrict our analysis of S1 and S3 to the region $m_a > E_b$, where $E_b$ denotes the ground-state binding energy, approximately $E_b \simeq \alpha_\chi^2 m_\chi/4$ in the Coulomb limit. In this region, bound-state formation through on-shell axion emission is kinematically forbidden at negligible incoming kinetic energy and thermally suppressed for $T \ll E_b$. We leave the region $m_a < E_b$ for future work.


\bigskip

\noindent{\it Acknowledgement} --- 
H.A. is supported by the National Science Foundation of China under Grant Nos. 12475107 and 12525506 and the National Key R\&D Program of China under Grant Nos. 2023YFA1607104 and 2021YFC2203100 and. 
F.G. is supported by the National Science Foundation of China under Grant No. 12521007, the Ministry of Education of China under Grant No. SRICSPYF-ZY2025028 and the Dushi program of Tsinghua University. 
J.L. is supported by the National Science Foundation of China under Grant Nos. 12235001 and 12475103, and State Key Laboratory of Nuclear Physics and Technology under Grant No. NPT2025ZX11.

OpenAI's ChatGPT (GPT-6.5 Astra) was used to independently cross-check selected analytical and numerical calculations presented in this work. The authors reviewed and independently verified the AI-assisted cross-checks, and all scientific conclusions were determined by the authors.
\bigskip

\appendix
\label{App:A}

\section{Dark-sector realizations and technical naturalness}
\label{app:UV}

We discuss renormalizable dark-sector realizations of the
four scenarios in Table~I, distinguishing protection
of the small fermion splitting from protection of the
light mediator mass. Scenarios S2 and S4 admit simple
pseudo-Nambu--Goldstone realizations. For S1 and S3,
the fermion spectrum and off-diagonal scalar coupling
can also be symmetry-protected, but obtaining an
unsuppressed coupling to a light ALP requires additional
structure.

\subsection{Pseudoscalar couplings: S2 and S4}

Introduce a complex SM-singlet scalar,
\begin{equation}
 \Phi=\frac{v_\phi+\rho}{\sqrt{2}}e^{ia/v_\phi},
 \;
 V(\Phi) \supset -\frac{b}{2}(\Phi^2+\Phi^{\dagger2}).
 \label{eq:UV_scalar}
\end{equation}
The global $U(1)_D$ is spontaneously broken by
$v_\phi$, while real $b>0$ softly breaks it explicitly.
For a real vacuum, the radial and ALP masses are
$m_\rho^2=2\lambda_\Phi v_\phi^2$ and $m_a^2=2b$.
In the isolated dark sector, $b\to0$ restores the
continuous symmetry and protects the light ALP.
Below, all displayed fermion parameters are real.

\paragraph{S2: Majorana fermions.}
Introduce left-handed Weyl fermions $\eta,\xi$ with
$U(1)_D$ charges $(q_\eta,q_\xi,q_\Phi)=(1,-1,-2)$.
The interactions are
\begin{equation}
 \mathcal L_2\supset
 -\frac{y}{2}\bigl(\Phi\eta\eta+
                  \Phi^\dagger\xi\xi\bigr)
 -\mu\eta\xi+\mathrm{h.c.}
 \label{eq:UV_S2}
\end{equation}
The equal Yukawa couplings follow from a unitary
internal exchange,
$E:\eta\leftrightarrow\xi,\ \Phi\leftrightarrow\Phi^\dagger$,
which is preserved by the scalar potential and vacuum.
Writing $M=y v_\phi/\sqrt{2}$, the mass matrix and its
eigenstates are
\begin{equation}
 \mathcal M=
 \begin{pmatrix}M&\mu\\ \mu&M\end{pmatrix},
 \qquad
 n_{1,2}=\frac{\eta\mp\xi}{\sqrt{2}}.
\end{equation}
For $0<\mu<M$,
\begin{equation}
 m_{1,2}=M\mp\mu,\qquad
 \Delta=2\mu,\qquad
 \mathcal L_a=-i\frac{M}{v_\phi}a\,n_1n_2
 +\mathrm{h.c.}
 \label{eq:UV_S2_match}
\end{equation}
In four-component Majorana notation this is
$(y/2)a\overline{\chi}_1i\gamma_5\chi_2+\mathrm{h.c.}$,
with $|y|=M/v_\phi$, up to an irrelevant overall sign.
Under $E$, $n_1$ and $a$ are odd while $n_2$ is even,
so diagonal single-ALP couplings are forbidden.

At $\mu=0$, the theory additionally preserves
independent fermion parities,
$Z_2^\eta:\eta\to-\eta$ and $Z_2^\xi:\xi\to-\xi$.
The bilinear $\eta\xi$ violates each separately.
Consequently, no additive mixing mass is generated
at any perturbative order.
In particular, the soft $\Phi^2+\mathrm{h.c.}$ term
preserves both parities and cannot generate $\mu$. 
Thus, $\Delta\simeq300$--$350,\mathrm{keV}$ requires $\mu\simeq150$--$175,\mathrm{keV}$, which is technically natural.

\paragraph{S4: Dirac fermions.}
Replace the Weyl pair by two Dirac fermions
$\Psi_A,\Psi_B$, with charges
\begin{equation}
 (q_\Phi,q_{AL},q_{AR},q_{BL},q_{BR})
 =(2,1,-1,-1,1).
\end{equation}
An exact common vector fermion number forbids
Majorana masses. Impose
$E:\Psi_A\leftrightarrow\Psi_B,\
\Phi\leftrightarrow\Phi^\dagger$, and take
\begin{equation}
\begin{aligned}
 \mathcal L_4\supset{}&
 -y\bigl(\Phi\overline{\Psi}_{AL}\Psi_{AR}
 +\Phi^\dagger\overline{\Psi}_{BL}\Psi_{BR}\bigr)\\
 &-\mu\bigl(\overline{\Psi}_{AL}\Psi_{BR}
 +\overline{\Psi}_{BL}\Psi_{AR}\bigr)
 +\mathrm{h.c.}
\end{aligned}
\end{equation}
The rotation $\chi_{1,2}=(\Psi_A\mp\Psi_B)/\sqrt{2}$
again gives $m_{1,2}=M\mp\mu$, and
\begin{equation}
 \mathcal L_a=-\frac{M}{v_\phi}a
 \bigl(\overline{\chi}_1i\gamma_5\chi_2+
       \overline{\chi}_2i\gamma_5\chi_1\bigr).
\end{equation}
This matches S4 with $|y|=M/v_\phi$.
At $\mu=0$, independent vector symmetries
$U(1)_A\times U(1)_B$ are restored, protecting
$\mu$ from additive corrections even for $b\neq0$.

In both constructions, protection of $\mu$ and
protection of equal diagonal masses are distinct.
If exchange is broken, a real symmetric mass matrix
with diagonal difference $\delta M$ gives
\begin{equation}
 \Delta=\sqrt{4\mu^2+\delta M^2}.
 \label{eq:UV_gap}
\end{equation}
The exchange symmetry is therefore essential to the
near-degeneracy and purely off-diagonal interaction.

\subsection{Scalar couplings: S1 and S3}

A simple realization uses a real mediator with an even
potential and $\langle a\rangle=0$. For S1, take
\begin{equation}
 \mathcal L_1\supset
 -\frac12(M-\epsilon)n_1n_1
 -\frac12(M+\epsilon)n_2n_2
 -g\,a\,n_1n_2+\mathrm{h.c.},
\end{equation}
where $n_i$ are left-handed Weyl fields. For S3, take
Dirac fields with an exact common fermion number:
\begin{equation}
\begin{aligned}
 \mathcal L_3\supset{}&
 -(M-\epsilon)\overline{\chi}_1\chi_1
 -(M+\epsilon)\overline{\chi}_2\chi_2\\
 &-g\,a\bigl(\overline{\chi}_1\chi_2+
             \overline{\chi}_2\chi_1\bigr).
\end{aligned}
\end{equation}
For real parameters, both match Table~I with
$|y|=|g|$ and $\Delta=2\epsilon$.

Two symmetries control these theories. The exact
internal sign transformation
\begin{equation}
 P:\quad a\to-a,\qquad
 \chi_1\to\chi_1,\qquad\chi_2\to-\chi_2
\end{equation}
(and analogously for $n_i$) forbids diagonal
single-$a$ couplings, constant off-diagonal masses. At $\epsilon=0$, the exchange
$E:\chi_1\leftrightarrow\chi_2,\ a\to a$ is restored, protecting the degeneracy.

\section{Constraints on heavy axion}
\label{sec:appendixB}

\begin{figure*}[ht!]
    \centering
    \includegraphics[width=1.7\columnwidth]{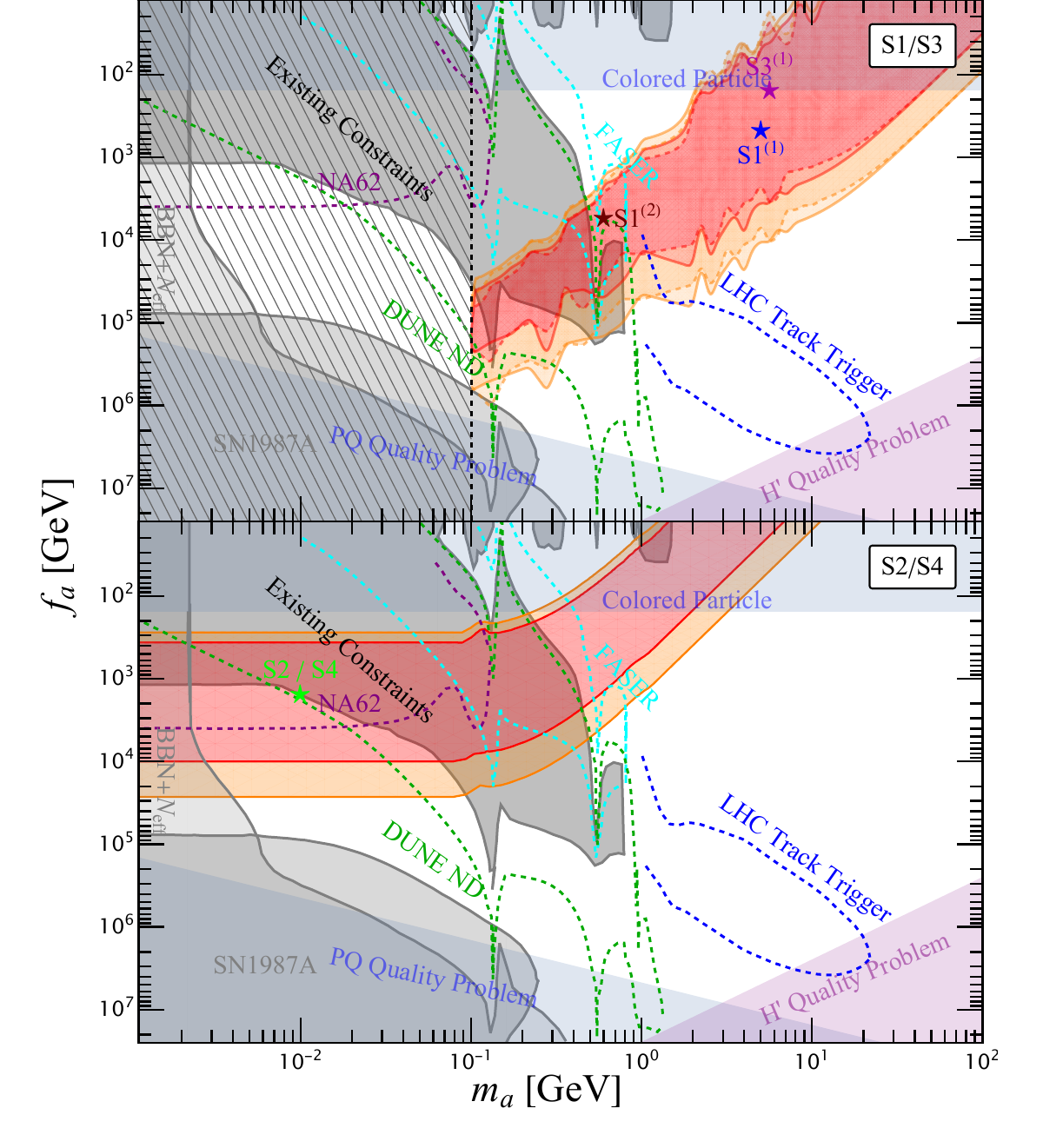}
    \caption{Experimental constraints and future sensitivities on the axion parameter space. The stars indicate the benchmark points used to calculate the recoil spectra shown in Fig.~\ref{fig:spec}. The red and orange shaded regions in the upper and lower panels denote the $1\sigma$ and $2\sigma$ regions, respectively, obtained from fitting our model to the LZ $248$~keV event. In the upper panel, the solid and dashed boundaries of the shaded regions correspond to the $\rm S1$ and $\rm S2$ scenarios, respectively, while in the lower panel the $\rm S2$ and $\rm S4$ regions overlap completely. The non-red and non-orange shaded regions show the experimental exclusion limits, while the dashed curves indicate future sensitivities, both taken from Ref.~\cite{Kelly:2020dda}. In the upper panel, the hatched region corresponds to $m_a < E_b$, where bound-state formation is possible. We leave the study of this region for future work.
}
    \label{fig:constraints}
\end{figure*}

Constraints on heavy axions with masses above $1\,\mathrm{MeV}$
have been extensively studied in the literature. These include
bounds from SN1987A~\cite{Ertas:2020xcc,Chang:2018rso},
cosmology~\cite{Depta:2020wmr}, and accelerator-based
searches~\cite{Dobrich:2015jyk,Dolan:2017osp,NA64:2020qwq,FASER:2018eoc,Gori:2020xvq,Aloni:2018vki,Mariotti:2017vtv}.
The existing constraints are summarized by the shaded regions
in Fig.~\ref{fig:constraints}. The dashed contours show the
projected sensitivities of future searches at
DUNE~\cite{Kelly:2020dda}, FASER~\cite{FASER:2018eoc},
the HL-LHC~\cite{Hook:2019qoh}, and NA62~\cite{Ertas:2020xcc}.

The benchmark points discussed in the main text are indicated
by stars in the same figure. Their locations relative to the
projected sensitivity contours illustrate the potential of
future experiments to probe the axion-portal inelastic
dark-matter scenario.

In Fig.~\ref{fig:constraints}, the red and orange shaded regions denote the $1\sigma$ and $2\sigma$ regions obtained by fitting our models to the LZ event, with S1 and S3 shown in the upper panel and S2 and S4 in the lower panel. As shown in the upper panels of Fig.~\ref{fig:spec}, explaining the LZ event in S1 and S3 requires $\alpha_\chi \sim \mathcal{O}(0.1)$, corresponding to a characteristic binding energy of $\alpha_\chi^2 m_\chi/4 \sim 0.1\,\mathrm{GeV}$ for the $\chi_{1,2}$ system. For $m_a \lesssim 0.1\,\mathrm{GeV}$, bound-state formation through mediator emission may therefore occur as the Universe cools, potentially modifying the viable parameter space of S1 and S3. A detailed treatment of this process is beyond the scope of this work. Our analysis therefore does not apply to the hatched region in the upper panel of Fig.~\ref{fig:constraints}.

\section{Non-perturbative calculation of dark-matter conversion}
\label{app:nonpert_conversion}

In the scalar-mediated scenarios S1 (Majorana) and S3 (Dirac),
the inelastic process $\chi_2\chi_2\to\chi_1\chi_1$ can be
substantially modified by repeated mediator exchange.  We describe
here how we calculate the velocity-dependent conversion cross section non-perturbatively.
The subsequent evolution of the excited-state
abundance, including the relation between the conversion rate and
$f_2$, is treated in the main text and is not repeated here.
We use $m_2-m_1\equiv\Delta>0$ and $m_1\simeq m_2\simeq m_\chi$.
The mediator has mass $m_a$ and off-diagonal scalar coupling $y$,
with $\alpha_\chi=y^2/(4\pi)$.  The condition $m_a>\Delta$
closes the two-body decay $\chi_2\to\chi_1 a$.

\subsection{Two-channel interaction and scattering states}
The non-perturbative effects of Yukawa-mediated dark matter interactions have been extensively investigated in previous studies, including inelastic Sommerfeld enhancement and resonant dark matter self-scattering~\cite{Slatyer:2009vg,Tulin:2012wi}. Here, we employ a coupled-channel partial-wave approach to calculate the non-perturbative conversion cross section for $\chi_2\chi_2\to\chi_1\chi_1$.
Each scalar-exchange vertex changes the dark-matter state, so the
exchange interaction connects the pair states $|11\rangle$ and
$|22\rangle$.  In the non-relativistic limit the leading potential is
\begin{equation}
\label{eq:appV}
\mathbf V(r)=\begin{pmatrix}
0&U(r)\\ U(r)&2\Delta
\end{pmatrix},\qquad
U(r)=-\frac{\alpha_\chi}{r}e^{-m_a r}.
\end{equation}
Here energies are measured relative to the $|11\rangle$ threshold;
the diagonal splitting is $2\Delta$, since both particles are
excited in the upper channel.  This potential is the central,
spin-independent scalar-exchange approximation used in our S1/S3
calculation.  Particle statistics are imposed when constructing the
physical partial-wave cross section below.

For an incident $|22\rangle$ pair of relative velocity $v$,
the reduced mass is $\mu\simeq m_\chi/2$ and
\begin{equation}
\label{eq:appk}
k_2=\mu v,\qquad
k_1=\sqrt{k_2^2+4\mu\Delta}
=\sqrt{k_2^2+2m_\chi\Delta}.
\end{equation}
The outgoing momentum thus remains finite as $v\to0$.
For each orbital angular momentum $\ell$, the reduced radial
wavefunctions $\mathbf u_\ell=(u_{1\ell},u_{2\ell})^T$ satisfy
\begin{equation}
\label{eq:appradial}
\mathbf u_\ell''(r)=\left[
\frac{\ell(\ell+1)}{r^2}\mathbf 1
+2\mu\mathbf V(r)-k_1^2\mathbf 1
\right]\mathbf u_\ell(r).
\end{equation}
Equivalently, the two diagonal equations have asymptotic wave
numbers $k_1$ and $k_2$ and are mixed only by $2\mu U(r)$.
This coupled problem resumes arbitrarily many exchanges of the
mediator within the potential-scattering approximation.

Regularity at the origin requires
$u_{i\ell}(r)\propto r^{\ell+1}$ for each independent regular
solution.  At sufficiently large radius, where $U(r)$ is
negligible compared with the channel kinetic energies, the
solution matrix can be expressed as a superposition of incoming
and outgoing free waves,
\begin{equation}
\label{eq:appasympt}
\mathbf u_\ell(r)\longrightarrow
\mathbf H_\ell^{(-)}(r)\,\mathbf a_{\rm in}
-\mathbf H_\ell^{(+)}(r)\,\mathbf a_{\rm out},\qquad
\mathbf a_{\rm out}=\mathbf S_\ell\mathbf a_{\rm in}.
\end{equation}
The diagonal matrices $\mathbf H_\ell^{(\pm)}$ contain
Riccati--Hankel functions evaluated at $k_i r$, normalized by the
appropriate channel flux factors.  This normalization is essential
for an inelastic reaction because $k_1\ne k_2$: it makes
$\mathbf S_\ell$ unitary when both channels are open and no
additional reaction channels are present.  With the ordering
$(|11\rangle,|22\rangle)$, the conversion probability for an
incident excited pair is
\begin{equation}
\label{eq:appP}
P_\ell(22\to11)=\left|S_{\ell,12}\right|^2,\qquad
0\le P_\ell\le1.
\end{equation}
In particular, $S_{\ell,12}$ is obtained from the \emph{off-diagonal}
scattering amplitude through numerically solving the Schr$\ddot{o}$dinger equation~\eqref{eq:appradial}.

\subsection{Partial waves and particle statistics}

The velocity-weighted conversion cross section is obtained by
summing the contributions of the allowed partial waves,
\begin{equation}
\label{eq:appsigv}
[\sigma v](v)=\frac{\pi}{\mu^2 v}
\sum_{\ell=0}^{\infty}(2\ell+1)\,w_\ell P_\ell(22\to11).
\end{equation}
The weights $w_\ell$ encode spin averaging and the identical-state
counting convention for the rate employed in the main text.
For S1, the two identical Majorana fermions must have an
antisymmetric total wavefunction; even orbital partial waves
occur in spin-singlet states and odd waves in spin-triplet states.
In our rate convention this gives
\begin{equation}
\label{eq:appweights}
w_\ell^{\rm S1}=\begin{cases}
\tfrac12,&\ell\ \text{even},\\
\tfrac32,&\ell\ \text{odd}.
\end{cases}
\end{equation}
For S3, the Dirac particle--particle, antiparticle--antiparticle,
and distinguishable particle--antiparticle configurations must be
counted with their corresponding symmetry and population factors.
We use the same convention in the conversion-rate and abundance
calculations, without applying the identical-particle factor twice.
The distinction between S1 and S3 therefore involves both their
couplings and their spin/statistical multiplicities.

A useful analytic reference is the low-velocity Born limit.  In
the S1 convention it reads
\begin{equation}
\label{eq:appborn}
[\sigma v]^{\rm S1}_{\rm Born}(0)=
\frac{y^4m_\chi^{3/2}\sqrt{\Delta}}
{2\sqrt{2}\,\pi\,(m_a^2+2m_\chi\Delta)^2}.
\end{equation}
For equal coupling $y$, the corresponding S3 rate in the
statistical convention used here has an additional factor $3/2$.
Notice that the denominator contains the nonzero outgoing
momentum scale $2m_\chi\Delta$, not merely $m_a^2$.
At large $\alpha_\chi m_\chi/m_a$, Born perturbation theory may
fail and individual perturbative partial-wave probabilities can
exceed unity.  Solving Eq.~\eqref{eq:appradial} yields a
unitary resummation, which can produce either enhancement or
suppression relative to Eq.~\eqref{eq:appborn}.


\bibliographystyle{elsarticle-num}
\bibliography{axion-elsarticle}



\null\clearpage

\end{document}